\documentclass{webofc}
\usepackage[varg]{txfonts}   
\begin{document}
\title{Recent results on cold-QCD from RHIC}
%
%

\author{\firstname{Qinghua} \lastname{Xu}\inst{1}\fnsep\thanks{\email{xuqh@sdu.edu.cn}}, for the STAR Collaboration 
}

\institute{
Institute of Frontier and Interdisciplinary Science \&  Key Laboratory of Particle Physics and Particle Irradiation of Minstry of Education,  
  Shandong University, Qingdao, Shandong, 266237, China
          }

\abstract{%
  Polarized proton-proton collisions at the Relativistic Heavy Ion Collider (RHIC) provide unique opportunities to study the spin structure of the nucleon. We will highlight recent results on the nucleon spin structure from the STAR and  PHENIX experiments at RHIC: (1) A sizable gluon polarization in the proton is measured with longitudinal double spin asymmetries of jet and hadron production; (2) Longitudinal single spin asymmetries in W boson production improve constraints on the sea quark polarization. The new spin asymmetry results for W boson confirmed the SU(2) flavor asymmetry of the light sea quark polarization in the proton; (3) Transverse spin effects in hadronic systems offer new implications on parton distribution functions in the collinear and transverse momentum dependent frameworks. We will also discuss near term plans for the STAR forward detector upgrade and prospects for proton-proton and proton-ion collisions in the years beyond 2021 at STAR.
}
\maketitle
\section{Introduction to RHIC spin program}
\label{intro}

In addition to colliding heavy ions, RHIC at Brookhaven National Laboratory is the world's only polarized proton-proton (pp) collider.  
By colliding high-energy beams of polarized protons, RHIC provides unique opportunities for exploring the spin structure of the proton~\cite{Bunce:2000uv}. 
In 1988 the EMC experimental results showed that the contribution from quark/antiquark spin to nucleon spin is surprisingly smaller than expected, leading to the so called ``spin puzzle".
The results from polarized Deep-Inelastic-Scattering (DIS) experiments in the past 30 years have shown that the spins of the quarks and antiquarks account for only about 30\% of proton's spin in the measured x-range.

Since 2001, RHIC has been providing pp collisions at center of mass energies of 200 GeV and 500 GeV with beams longitudinally or transversely polarized. 
There have been two main detectors, PHENIX and STAR, at RHIC to perform spin experiments as well as heavy ion ones. PHENIX completed its data taking in 2016. STAR is currently the only experiment running at RHIC.


In the following, we will highlight recent results on nucleon spin structure from the STAR and PHENIX experiments.
RHIC spin experiments have been providing information on how much gluon spin contributes to the spin of the proton, which is observed to be sizable.
Significant constraints on the sea quark polarization are also obtained beyond the semi-inclusive DIS measurements, in particular the SU(2) flavor asymmetry of the light sea quark polarization is confirmed through $W$ boson production. 
Information on how quarks move around inside the proton can be gained by studying transverse spin effects at RHIC, which offer new implications on parton distribution functions in the collinear and transverse momentum dependent (TMD) frameworks. 
Finally we will discuss future plans for the STAR forward detector upgrade and prospects for pp and proton-ion (pA) collisions in the years beyond 2021 at RHIC.

\section{Recent highlights from RHIC spin}
\label{sec-1}
%
\subsection{Gluon polarization determination with jet/hadron production }
\label{sub1}
\begin{figure}[b]
\centering
\vspace*{-0.65cm}       
\includegraphics[width=7.8cm,clip]{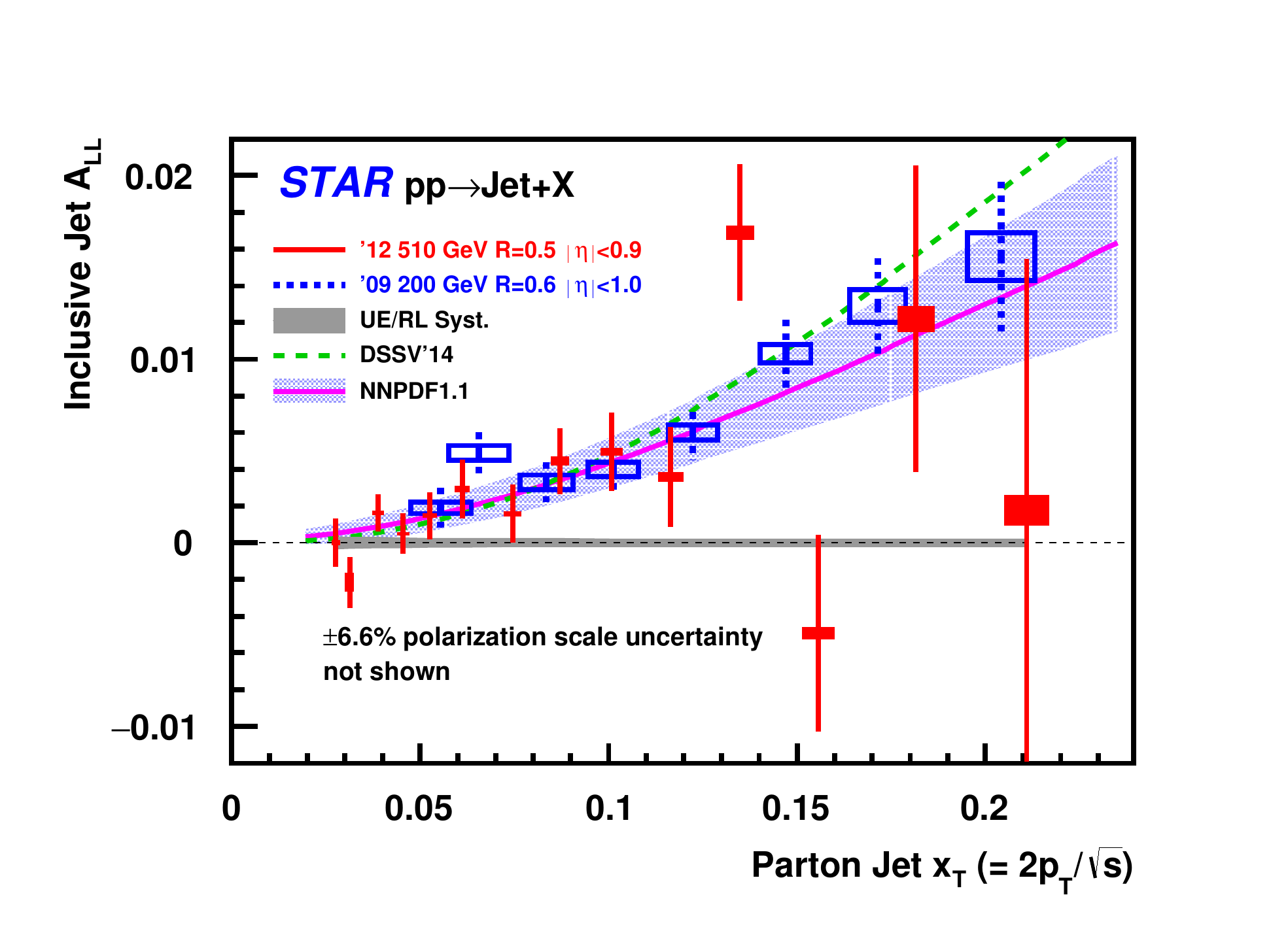} 
\includegraphics[width=5.8cm,clip]{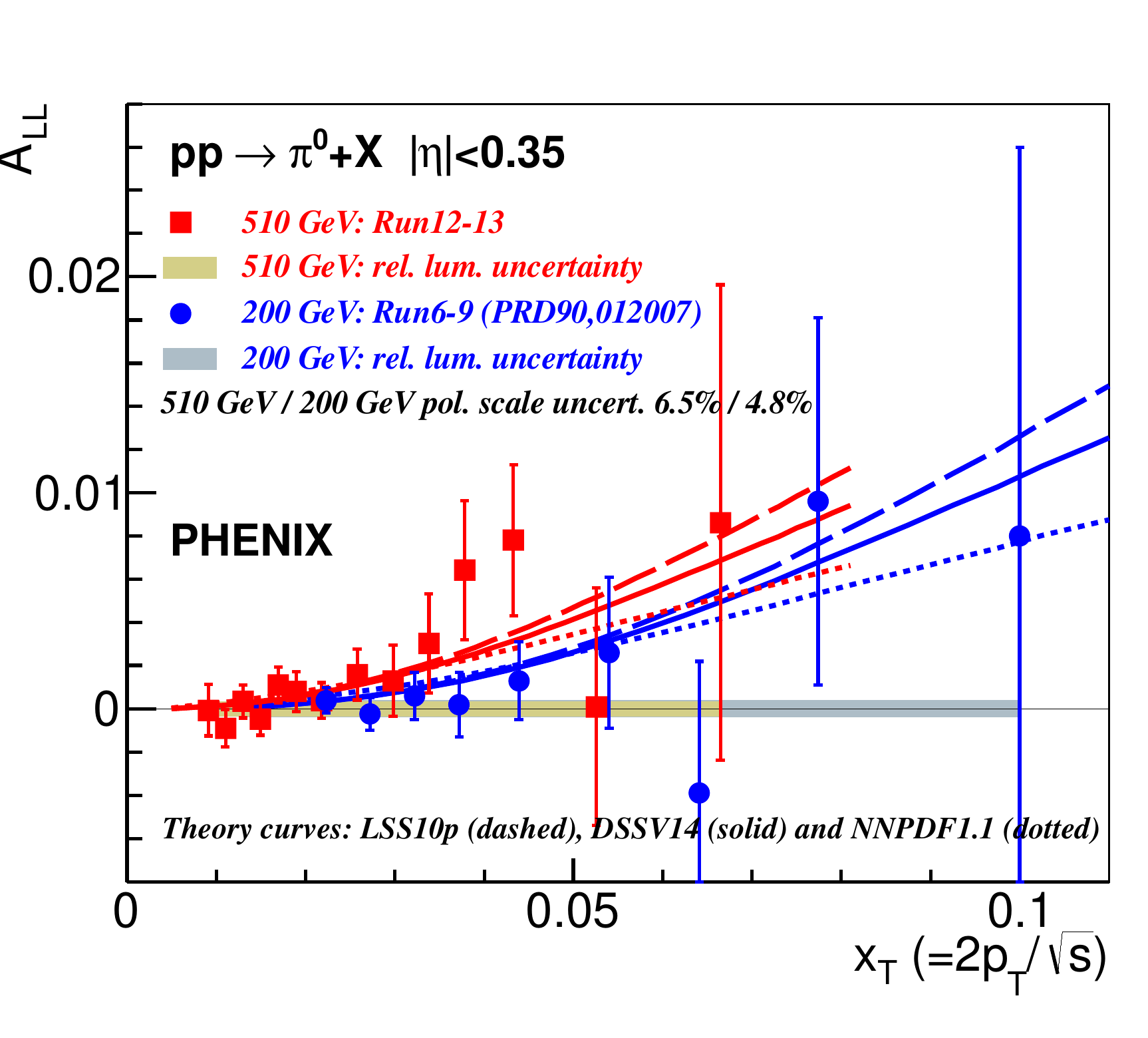}
\vspace*{-0.5cm}
\caption{(left) Results of longitudinal double spin asymmetry $A_{LL}$ for inclusive jet production in pp collisions at STAR at 200 GeV and 510 GeV~\cite{Adam:2019aml}. 
(right) Results of longitudinal double spin asymmetry $A_{LL}$ for $\pi^0$ production in pp collisions at PHENIX at 200 GeV and 510 GeV~\cite{Adare:2015ozj}. 
}
\label{figure1}       
\end{figure}

The gluon polarization was once expected to account for most of the "missing" part of the proton spin. 
In lepton-nucleon DIS process, it can be determined through scaling violations and gluon splitting process, but was limited by kinematic range and statistics.
In pp collisions at RHIC, the gluon helicity distribution $\Delta g(x)$ can be accessed with strongly interacting probes, in jet or hadron production, by measuring the longitudinal double-spin asymmetries ($A_{LL}$),
\begin{equation}
A_{LL} =\frac{(\sigma^{++}+\sigma^{--})-(\sigma^{+-}+\sigma^{-+})}{(\sigma^{++}+\sigma^{--})+(\sigma^{+-}+\sigma^{-+})}
= \frac{1}{P_1 P_2}\frac{(N^{++}+N^{--}) - R (N^{+-}+N^{-+})}{(N^{++}+N^{--}) + R (N^{+-}+N^{-+})},
\end{equation}
where $\sigma$ denotes the cross section for jet/hadron production and the superscripts "+" or "-" denote the helicities of the two proton beams.  $P_1$ and $P_2$ are the beam polarizations, $R=(\mathcal{L}^{++}+\mathcal{L}^{--})/(\mathcal{L}^{+-}+\mathcal{L}^{-+})$
is the relative luminosity ratio, and $N$'s are the corresponding jet/hadron yields.
%
In QCD, assuming factorization, $A_{LL}$ is proportional to the convolution of parton helicity distributions, the double spin asymmetry for the hard partonic scattering, and the fragmentation function for hadron production.  A significant fraction of the partonic scatterings is gluon-involved such as gluon-gluon or quark-gluon scattering for jet and hadron production at RHIC, which thus provide sensitivity to the gluon helicity distribution function~\cite{Bunce:2000uv}.

\begin{figure*}
\centering
\includegraphics[width=4.5cm,clip]{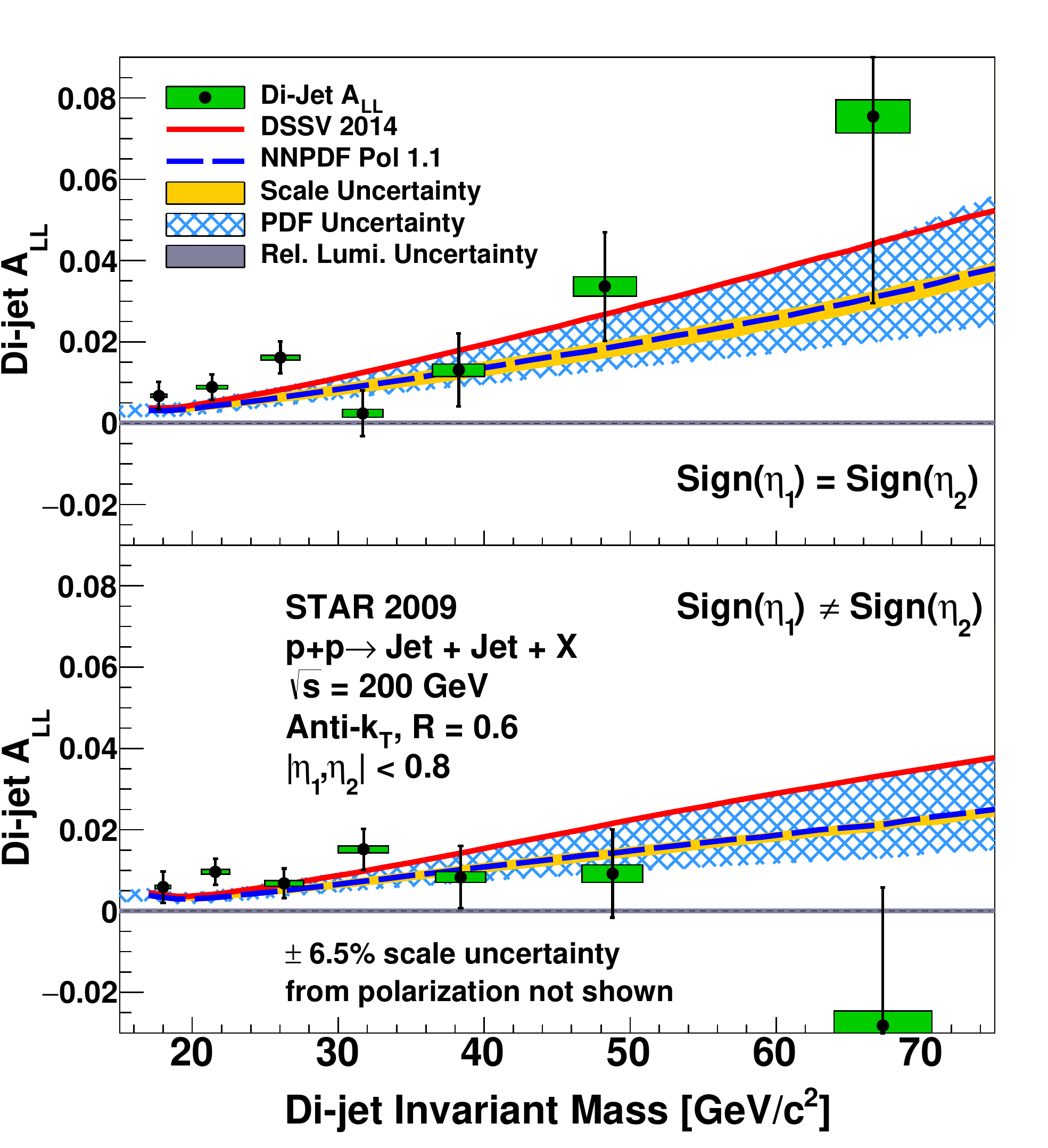} 
\includegraphics[width=5.1cm,clip]{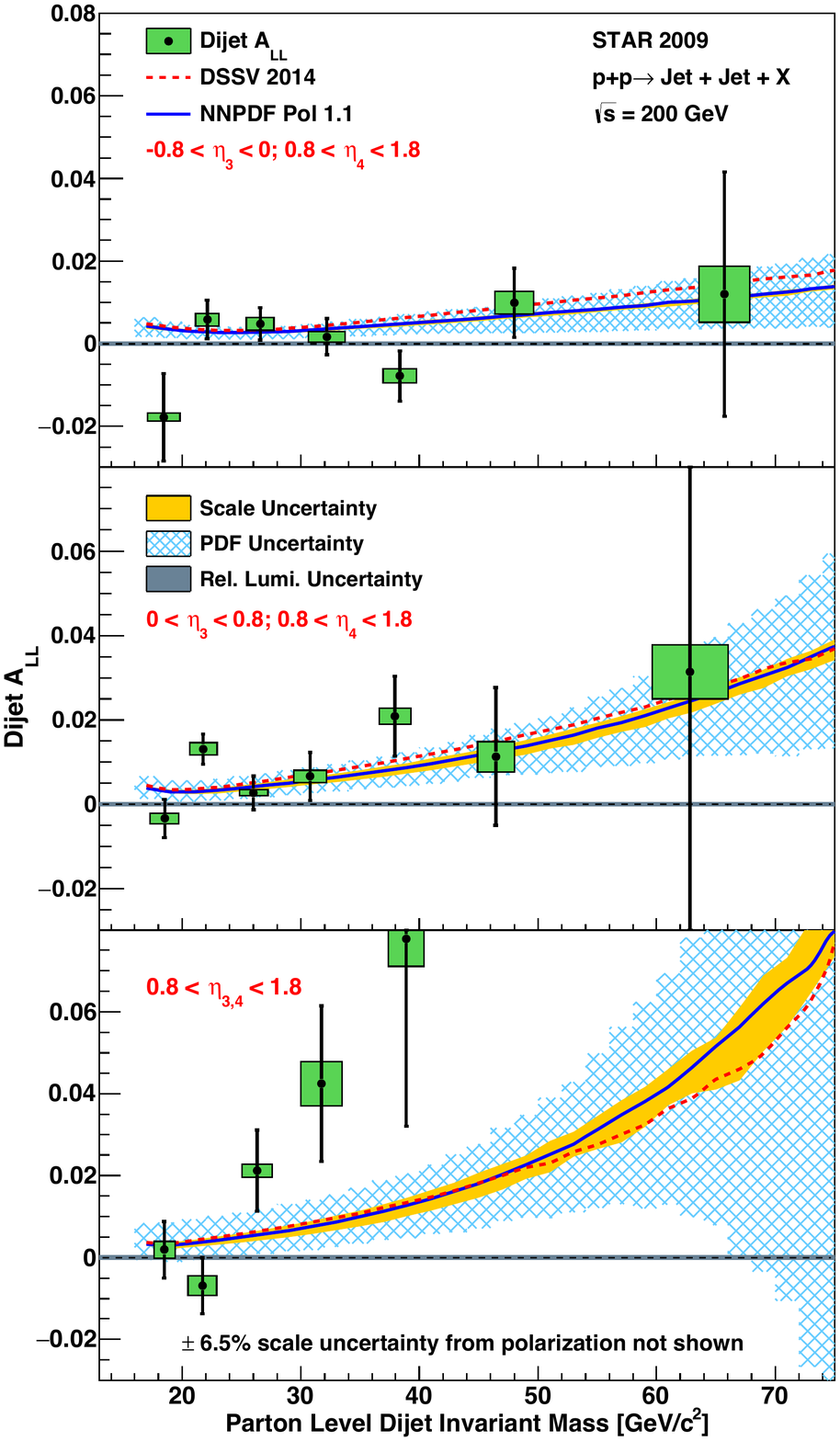}
\includegraphics[width=4.37cm,clip]{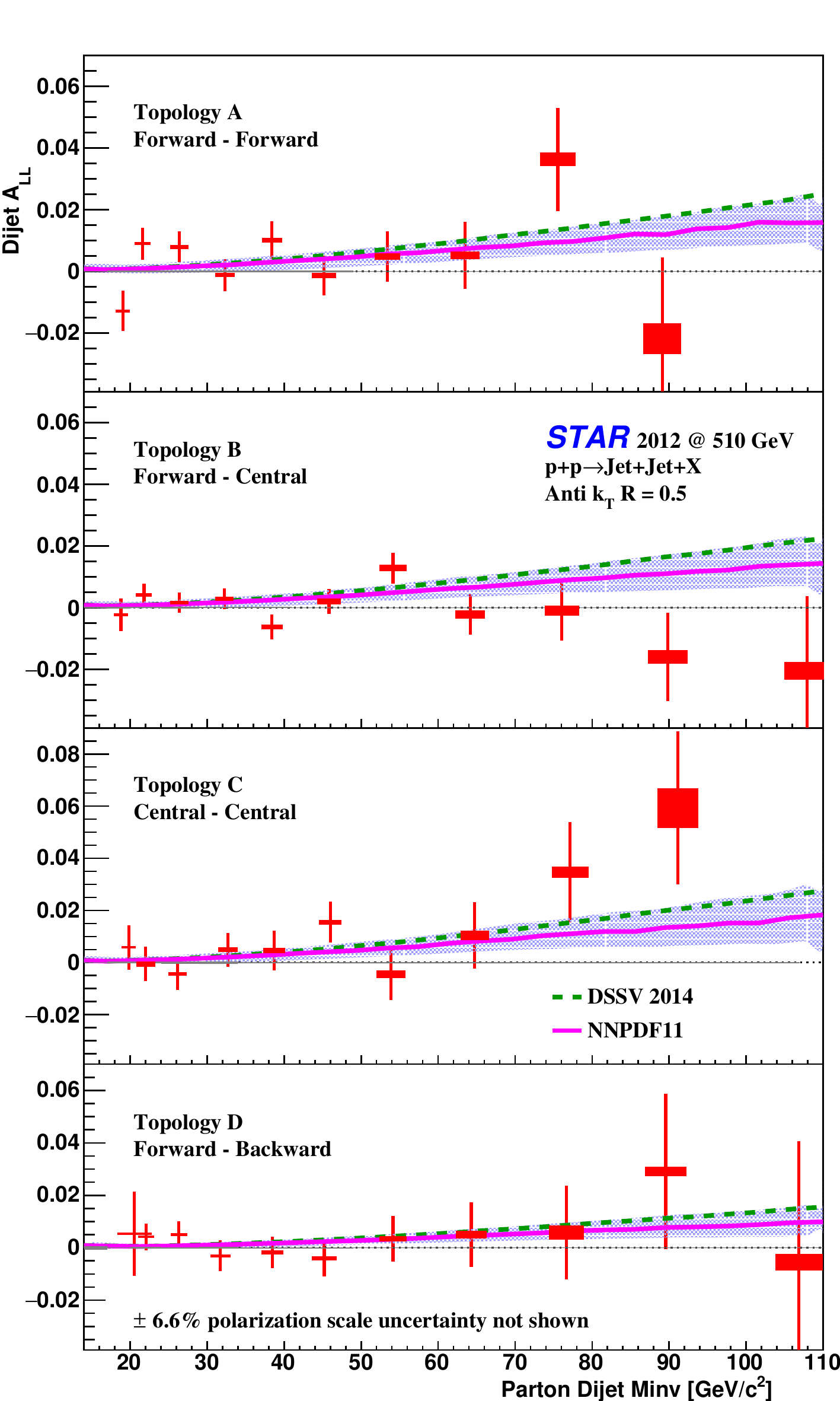} 
\caption{(left) $A_{LL}$ results for di-jet production in mid-rapidity in pp collisions at STAR at 200 GeV~\cite{Adamczyk:2016okk}.
(middle)  $A_{LL}$ results for di-jet production in mid \& near forward rapidity in pp collisions at STAR at 200 GeV~\cite{Adam:2018pns}.
(right) $A_{LL}$ results for di-jet production in mid-rapidity in pp collisions at STAR at 510 GeV~\cite{Adam:2019aml}.
}
\label{figure2}       
\vspace*{-0.5cm} 
\end{figure*}

The first evidence of non-zero gluon polarization has been provided by $A_{LL}$ measurements for inclusive jets at STAR~\cite{Adamczyk:2014ozi} and for $\pi^0$ at PHENIX~\cite{Adare:2014hsq} in pp collisions at 200 GeV. 
The global analyses from the DSSV and NNPDF groups indicated that the integral of the gluon helicity distribution within the kinematic range of $0.05<x<0.2$ could be as large as about 0.2~\cite{deFlorian:2014yva, Nocera:2014gqa}. 
To probe the gluon polarization in the lower $x$ region, STAR recently published $A_{LL}$ results for inclusive jets at 510 GeV~\cite{Adam:2019aml}, and PHENIX also published their $A_{LL}$ results for inclusive $\pi^0$ at 510 GeV~\cite{Adare:2015ozj}. 
Both results are shown in Fig.~\ref{figure1}, which can provide constraints on gluon polarization with $x$ down to 0.015. 
In addition, di-jet correlation measurements can provide information on the $x$-dependence of the gluon helicity distribution.
Figure~\ref{figure2} shows the $A_{LL}$ results for di-jet measurements with STAR at 200 GeV and 510 GeV with different detector coverages~\cite{Adamczyk:2016okk,Adam:2018pns,Adam:2019aml}. The recent STAR di-jet results have been included in a global analysis and indicated their significance in constraining the shape of $\Delta g(x)$~\cite{deFlorian:2019zkl}.
%
%

\subsection{Probing sea quark polarization via $W$ boson production}
\label{sub2}
The spin contribution of the sea quarks is also an important piece to the complete understanding of the nucleon spin structure.  The production of $W^\pm$ bosons in pp collisions with one beam longitudinally polarized provides an unique clean probe of the sea quark helicity distributions without the complication of hadron fragmentation, as in semi-inclusive DIS~\cite{Bunce:2000uv}.
At RHIC, the longitudinal single-spin asymmetry $A_L$ for $W$ boson production in pp collisions is defined as:
$A_L\equiv ({\sigma^+ - \sigma^-})/({\sigma^+ + \sigma^-})$,
where $\sigma^{+/-}$ is the $W$ cross section with a positive/negative longitudinal spin orientation of the proton beam.
At RHIC, the $W$'s are detected via leptonic decay, which are characterized by an isolated $e^{\pm}$ or $\mu^{\pm}$ with a sizable transverse energy, which peaks near half the $W$ mass.

\begin{figure*}
\centering
\sidecaption
\vspace*{-0.3cm}
\includegraphics[width=5.1cm,clip]{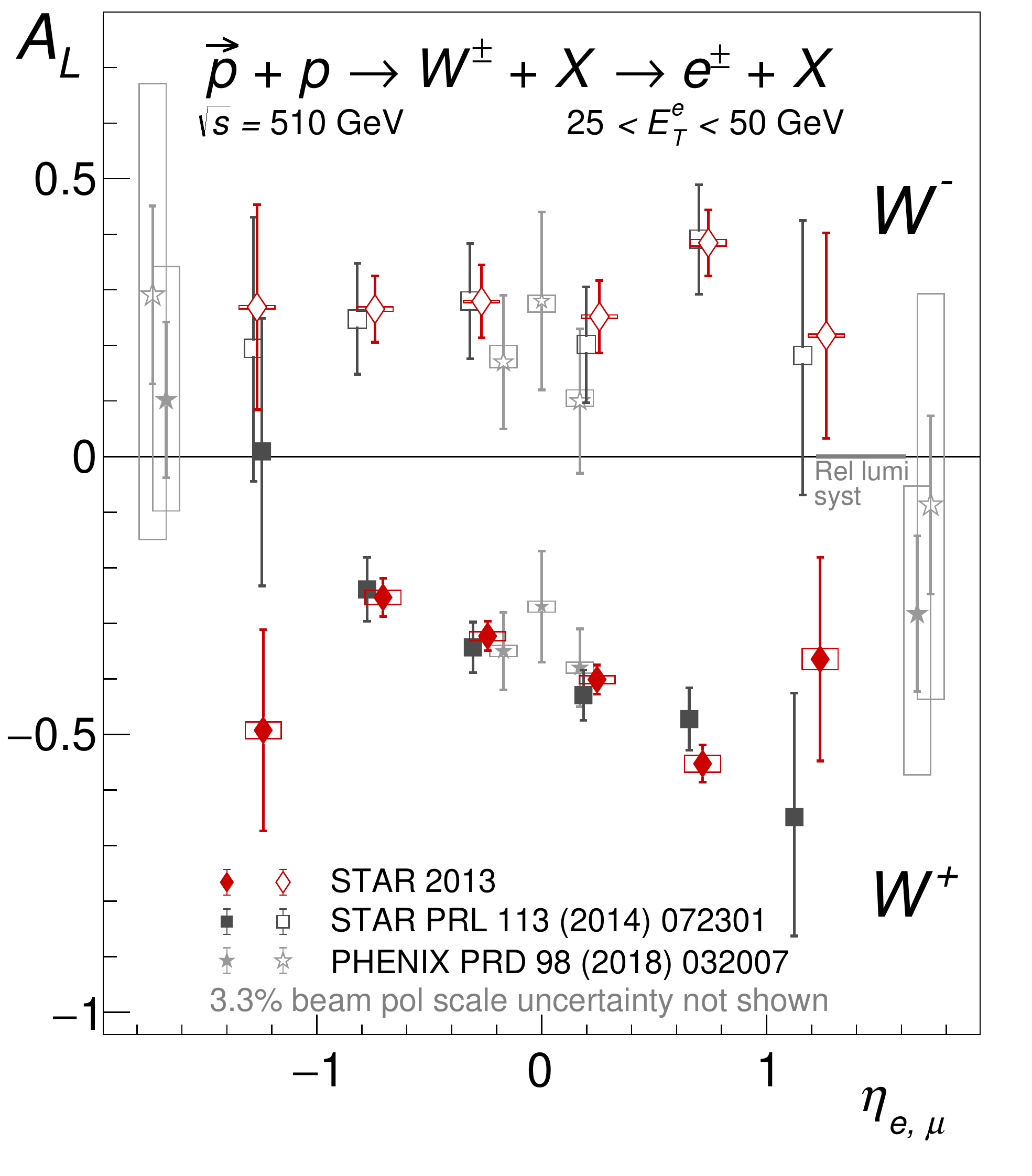} 
\includegraphics[width=5.1cm,clip]{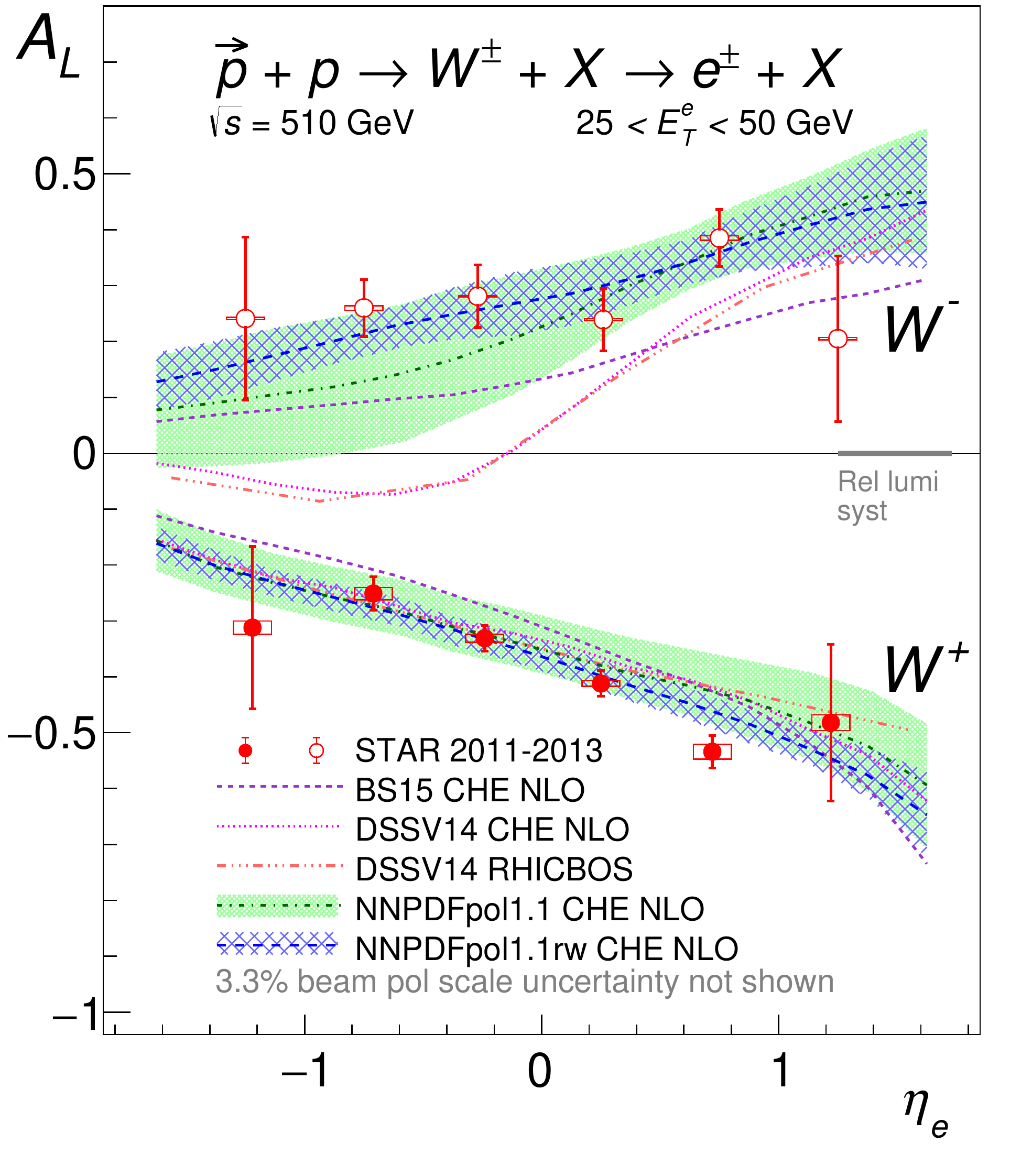}
\caption{(left)Results of longitudinal single spin asymmetry $A_{L}$ for $W$ boson from STAR~\cite{STAR13} and PHENIX~\cite{Adare:2015gsd,Adare:2018csm} from pp data at 510 GeV. 
(right) Combined results on $A_{L}$ for $W$ boson from STAR data 2011-2013~\cite{STAR13}, and compared with theoretical calculations. 
}
\label{figure3}       
\vspace*{-0.7cm}
\end{figure*}

The final $A_L$ results for $W^\pm$ boson versus lepton pseudorapidity from the STAR 2013 data sample \cite{STAR13} are shown in the left panel of Fig. \ref{figure3}, in comparison with STAR results from the  2011+2012 data \cite{Adamczyk:2014xyw} and the final PHENIX results of $A_L$ for leptons from $W/Z$ decay~\cite{Adare:2015gsd,Adare:2018csm}. 
The 2013 data sample corresponds to an integrated luminosity of about 250 pb$^{-1}$ with an average beam polarization of about 56\%. This is about 3 times the previous data sample taken in 2011 and 2012 \cite{Adamczyk:2014xyw}, and thus the most precise measurement of $W$ $A_L$ at RHIC. 
The combined STAR data from years 2011-2013 are shown in the right panel of Fig.~\ref{figure3}, in comparison with theoretical expectations based on the different inputs of polarized parton densities.

\begin{figure*}[b]
\centering
\vspace*{-0.3cm}
\includegraphics[width=4.15cm,clip]{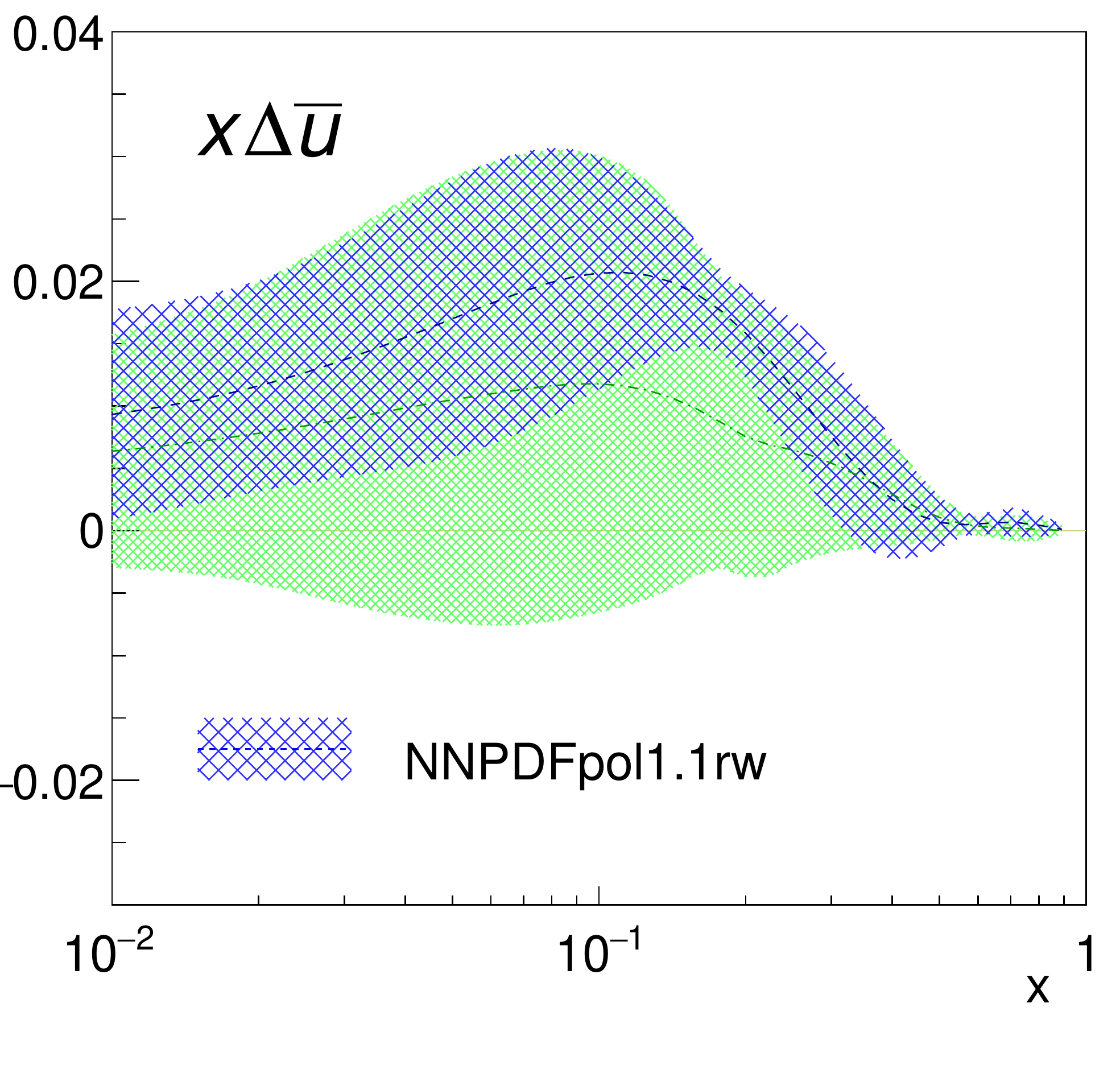} 
\includegraphics[width=4.2cm,clip]{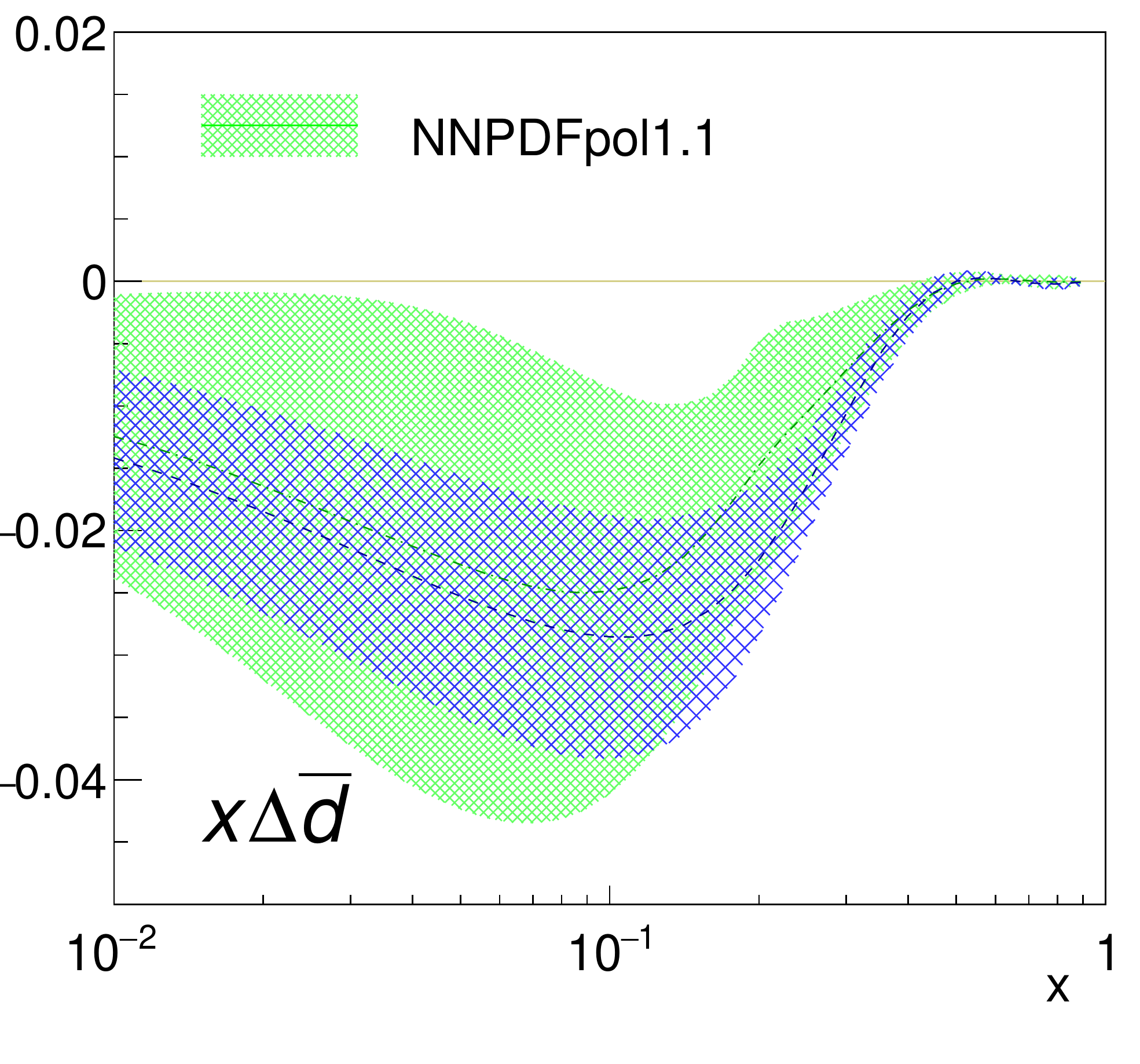}
\includegraphics[width=4.1cm,clip]{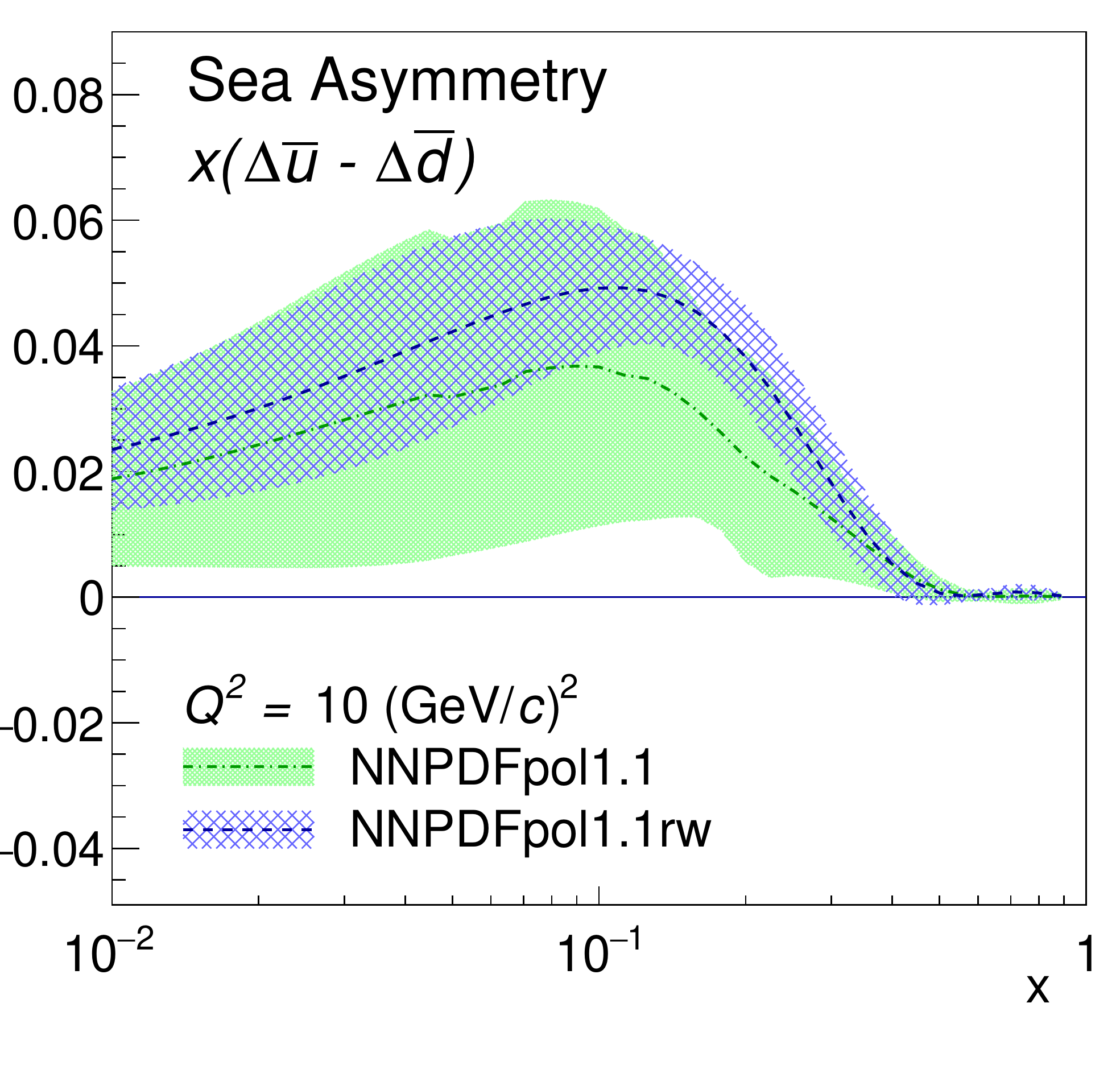} 
\vspace*{-0.5cm}
\caption{
The helicity distributions of $\bar u$ and $\bar d$ quarks and their difference $\Delta\bar{u}-\Delta\bar{d}$ as a function of $x$ at a scale of $Q^2$ = 10\,(GeV/$c$)$^2$.  
The green curve/band shows the NNPDFpol1.1 result~\cite{Nocera:2014gqa} and the blue curve/band shows the corresponding distribution after the 2013 $W^\pm$ data are included by reweighting~\cite{STAR13}.
}
\label{figure4}       
\vspace*{-0.7cm}
\end{figure*}

To assess their impact in constraining the sea quark helicity distribution, the STAR 2013 $A_L$ data were used in the reweighting procedure of NNPDF global fit based on NNPDFpol1.1 parton densities~\cite{Nocera:2014gqa}. 
The results from the reweighting are shown in Fig.~\ref{figure3} (right plot, blue hatched bands), and the uncertainties are significantly reduced compared to before the reweighting.
The helicity distributions of $\bar u$ and $\bar d$ quarks in the proton and their difference $\Delta\bar{u}(x)-\Delta\bar{d}(x)$ from the reweighting are shown in Fig.~\ref{figure4}~\cite{STAR13}.
The new results confirm the existence of a flavor asymmetry in the polarized quark sea, $\Delta\bar{u}(x)>0>\Delta\bar{d}(x)$, in the range of 0.05 $< x <$ 0.25 at a scale of $Q^2 = 10$\,(GeV/$c$)$^2$.
This is opposite to the flavor asymmetry observed in the unpolarized quark distributions, where $\bar{d}(x)>\bar{u}(x)$ over a wide $x$ range has been observed in the Drell-Yan process~\cite{Towell:2001nh}. 

\subsection{Transverse spin physics results}
\label{sub3}

Significant transverse single spin asymmetries ($A_N$) have been observed for different hadron production in hadron-hadron collisions over a wide range of colliding energies since the 1970's.
STAR measurements have demonstrated the persistence of sizeable $A_N$ for forward $\pi^0$ production at RHIC energies~\cite{Abelev:2008af}, where different mechanisms including the higher twist effect, and TMD effects like the Sivers or Collins fragmentation effect could all contribute. 
Therefore, it is important to study different effects separately for a full understanding of the underlying mechanism. 

\begin{figure}[h]
\vspace{-0.35cm}
\centering
\sidecaption
\includegraphics[width=8.9cm,clip]{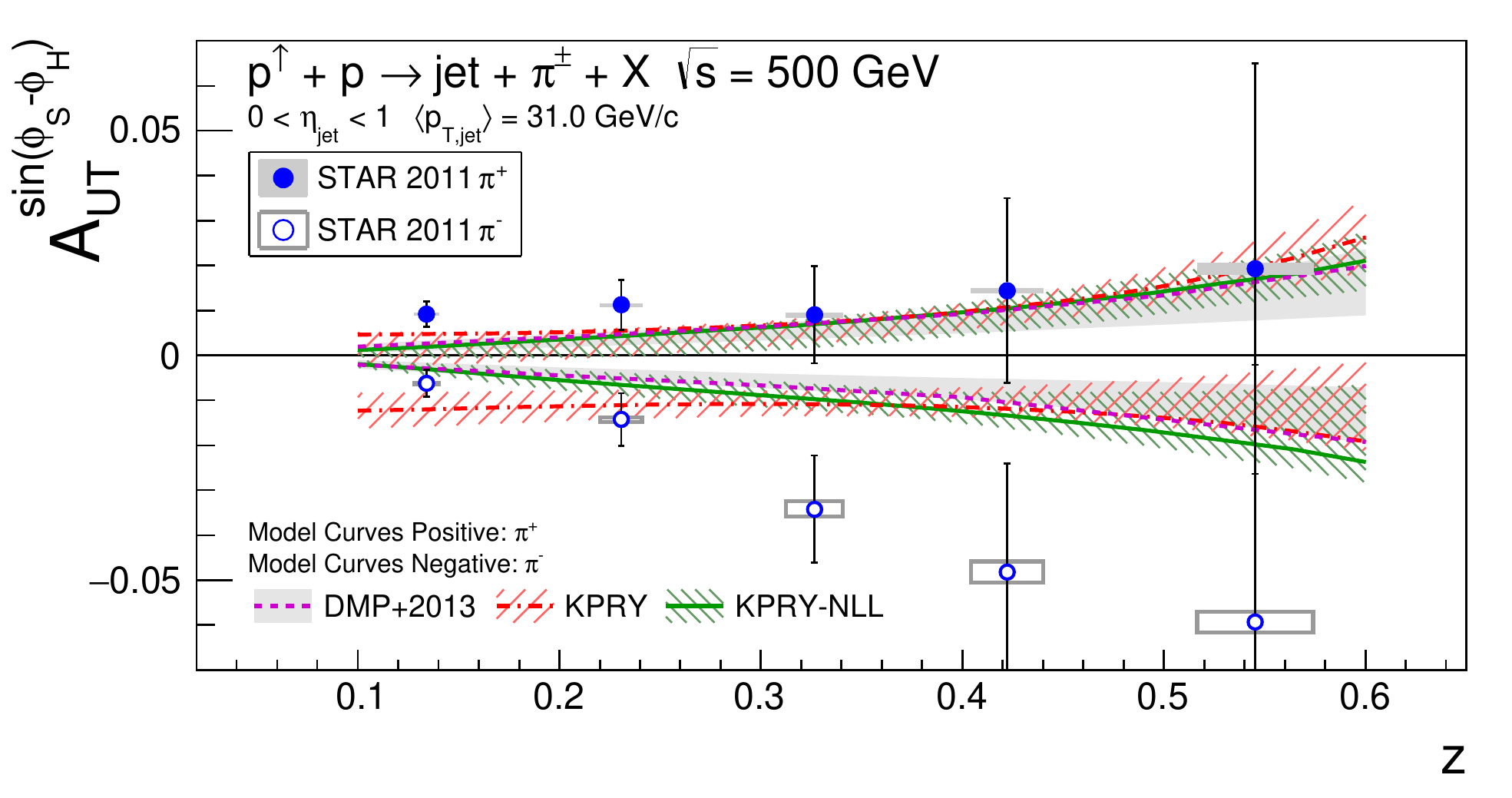}
\caption{
Collins asymmetries as a function of pion $z$ for jets reconstructed with $<p_T>$=31 GeV and 0 < $\eta$ < 1 in pp collisions at 500 GeV at STAR~\cite{Adamczyk:2017wld}. The asymmetries are shown in comparison with model calculations.
}
\vspace{-0.75cm}
\label{figure5}       
\end{figure}

\begin{figure}[h]
\vspace{-0.3cm}
\centering
\sidecaption
\includegraphics[width=7.5cm,clip]{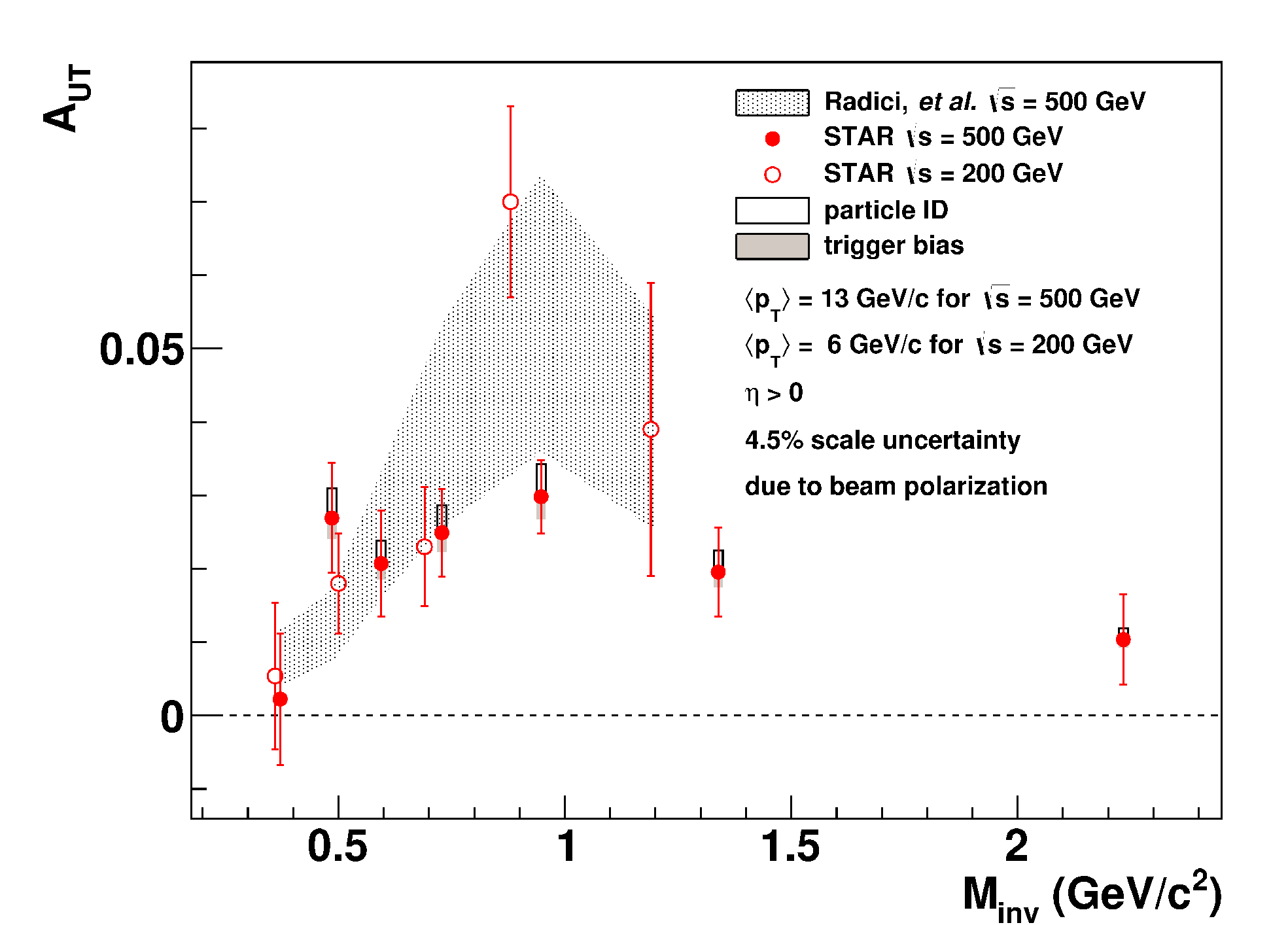}
\caption{The azimuthal asymmetry $A_{UT}$ for $\pi^+ \pi^-$  pairs as a function of invariant mass at 200 GeV and 500 GeV~\cite{Adamczyk:2017ynk}, and compared with predictions.}
\vspace{-0.65cm}
\label{figure6}       
\end{figure}

The hadron production within a jet in pp collisions provides direct access to the Collins fragmentation
function, and the corresponding $A_N$ is also connected to proton transversity. 
Figure \ref{figure5} shows the results of the mid-rapidity Collins asymmetries as a function of $z$ for
charged pions in pp collisions at 500 GeV at STAR~\cite{Adamczyk:2017wld}, where $z$ is the momentum fraction of the jet carried by the pion.
STAR also made measurements of transverse single spin asymmetries in di-hadron production at mid-rapidity~\cite{Adamczyk:2017ynk},
which provide another channel to access proton transversity through the analyzing power of the
interference fragmentation functions (IFF).
Figure \ref{figure6} shows the results for the IFF asymmetry $A^{sin\phi}_{UT}$
 at mid-rapidity as a function of di-hadron invariant mass at 200 GeV and 500 GeV. 

The transverse single spin asymmetry for $W$ and $Z$ production in pp collisions provides an excellent opportunity to test the sign-change of the Sivers function, in comparison to semi-inclusive DIS. STAR published the first results on $W$ boson $A_N$~\cite{Adamczyk:2015gyk}, which indicated a preference for the sign-change.
The analysis of a 14 times larger data sample taken in 2017 at STAR is underway, and will provide an important test of the sign-change of the Sivers function.

The possible nucleus dependence of single spin asymmetry has been predicted and also measured at RHIC.
PHENIX measurements of the single spin asymmetry of forward neutron in proton-nucleus collisions indicated a possible A-dependence~\cite{Aidala:2019ctp}, while the STAR measurements of forward $\pi^0$ do not show this trend~\cite{Steven}. In addition, STAR made the first measurement of the transverse spin transfer for $\Lambda$ and $\bar\Lambda$ hyperons in transversely polarized pp collisions at 200 GeV~\cite{Adam:2018wce}, which may provide insights into the transversity distribution of the strange quark.

\section{Summary and future plan for cold-QCD physics at RHIC}
\label{summary}

RHIC continues its efforts to deepen our understanding of the nucleon spin structure.
Sizable gluon polarization in the proton has been determined from RHIC measurements with polarized proton beams.
New results on double spin asymmetries $A_{LL}$ for inclusive jets, di-jets and $\pi^0$ production from the STAR and PHENIX experiments in pp collisions at 200 and 510 GeV, are providing further constraints on the gluon helicity distribution.
Both PHENIX and STAR experiments also published their final results on the single-spin asymmetry $A_L$ for $W$ boson production from the largest data sample taken in 2013. 
The new $A_L$  results from STAR further confirmed the SU(2) flavor asymmetry in sea quark polarization, i.e.,  $\Delta\bar{u}(x) >0> \Delta\bar{d}(x)$ in the range 0.05 $<x<$ 0.25 at $Q^2 \,\sim\,10$\,(GeV/$c$)$^2$.
On the transverse spin program, new results on single spin asymmetries of the Collins effect for pion production within a jet at mid-rapidity are reported from STAR.
STAR also made measurements of di-hadron spin asymmetries (IFF),  which provide insights into the proton transversity distribution.
The transverse single spin asymmetry for $W$ and $Z$ production with a large data sample taken in 2017 at STAR will provide a great opportunity to test the sign-change of the Sivers function.

STAR is currently performing a detector upgrade in the forward rapidity region 2.5<$\eta$<4, which includes a Forward Tracking System (FTS) and a Forward Calorimeter System (FCS). 
The FTS consists of 3 layers of silicon mini-strip disks and 4 layers of small-strip Thin Gap Chamber, and the FCS includes both electro-magnetic and hadron calorimeters. 
The upgrade will be completed in late 2021 and will provide improved calorimetry, tracking, and charge identification and photon hadron separation in the forward region, which will enable the measurements of full jets, Drell-Yan, and prompt photons in
pp and pA collisions.
Dedicated polarized pp running is expected at RHIC-STAR in the year of 2021/2022, and then STAR will continue to run pp/pA/AA in parallel with sPHENIX beyond 2021. 
In particular, the proposed pp/pA measurements will be essential to fully realize the scientific promise of the EIC collider~\cite{coldQCD2016}. 

The author is supported partially by the National Natural Science Foundation of China (No. 11520101004).

\end{document}